\documentclass[12pt,preprint]{aastex}
\usepackage{amsmath,longtable,natbib,upgreek}
\usepackage{lineno}
\usepackage{hyperref}
\shorttitle{Transition MSP J1227-4853}
\shortauthors{Roy et al.}

\begin{document}

\title{Discovery of PSR J1227$-$4853: A transition from a low-mass X-ray binary to a redback millisecond pulsar}

\author{
Jayanta~Roy\altaffilmark{1,2},
Paul~S.~Ray\altaffilmark{3},
Bhaswati~Bhattacharyya\altaffilmark{1},
Ben~Stappers\altaffilmark{1},
Jayaram~N.~Chengalur\altaffilmark{2},
Julia~Deneva\altaffilmark{4},
Fernando~Camilo\altaffilmark{5},
Tyrel~J.~Johnson\altaffilmark{6},
Michael~Wolff\altaffilmark{3},
Jason~W.~T.~Hessels\altaffilmark{7,8},
Cees~G.~Bassa\altaffilmark{7},
Evan~F.~Keane\altaffilmark{9}, 
Elizabeth~C.~Ferrara\altaffilmark{10},
Alice~K.~Harding\altaffilmark{10}
Kent~S.~Wood\altaffilmark{3}
}
\altaffiltext{1}{Jodrell Bank Centre for Astrophysics, University of Manchester, M13 9PL,  UK}
\altaffiltext{2}{National Centre for Radio Astrophysics, Tata Institute of Fundamental Research, Pune 411 007, India}
\altaffiltext{3}{Space Science Division, Naval Research Laboratory, Washington, DC 20375-5352, USA}
\altaffiltext{4}{NRC Research Associate, resident at Naval Research Laboratory, Washington, DC 20375-5352, USA}
\altaffiltext{5}{Columbia Astrophysics Laboratory, Columbia University, New York, NY 10027, USA}
\altaffiltext{6}{College of Science, George Mason University, Fairfax, VA 22030, USA, resident at Naval Research Laboratory, Washington, DC 20375, USA}
\altaffiltext{7}{ASTRON, the Netherlands Institute for Radio Astronomy, Postbus 2, 7990 AA, Dwingeloo, The Netherlands}
\altaffiltext{8}{Anton Pannekoek Institute for Astronomy, University of Amsterdam, Science Park 904, 1098 XH Amsterdam, The Netherlands}
\altaffiltext{9}{Centre for Astrophysics and Supercomputing, Swinburne University of Technology, Mail H30, PO Box 218, VIC 3122, Australia}
\altaffiltext{10}{NASA Goddard Space Flight Center, Greenbelt, MD 20771, USA}
\affil{}

\begin{abstract} 
XSS J12270$-$4859 is an X-ray binary associated with the \textit{Fermi} LAT gamma-ray source 1FGL J1227.9$-$4852. In 2012 December, this source underwent a transition 
where the X-ray and optical luminosity dropped and the spectral signatures of an accretion disc disappeared. We report the discovery of a 
1.69 millisecond pulsar (MSP), PSR J1227$-$4853, at a dispersion measure of 43.4 pc cm$^{-3}$ associated with this source, using the GMRT at 607\,MHz. 
This demonstrates that, post-transition, the system hosts an active radio MSP.  This is the third system after PSR J1023$+$0038 and PSR J1824$-$2452I 
showing evidence of state switching 
between radio MSP and low-mass X-ray binary (LMXB) states. We report timing observations of PSR J1227$-$4853 with the GMRT and Parkes, which give a precise 
determination of the rotational and orbital parameters of the system. 
The companion mass measurement of 0.17 to 0.46 M$_{\odot}$ suggests that this is a redback 
system. PSR J1227$-$4853 is eclipsed for about 40\% of its orbit at 607\,MHz; with additional short-duration eclipses at all orbital phases. We also find that the pulsar is very energetic, 
with a spin-down luminosity of $\sim$ 10$^{35}$ erg s$^{-1}$. We report simultaneous imaging and timing observations with the GMRT, 
which suggests that eclipses are caused by absorption, rather than dispersion smearing or scattering.  
\end{abstract}

\vskip 0.6 cm

\keywords{pulsars: general; binaries: eclipsing, pulsars: individual (PSR J1227$-$4853)}
\section{Introduction}
\label{sec:intro}
Neutron star low-mass X-ray binaries (LMXBs) and radio millisecond pulsars (MSPs) are evolutionarily linked, where MSPs are the end products 
of an episode of accretion of matter and angular momentum from the binary companion to the neutron star \citep{Bhattacharya91}. Recent observations of 
PSR J1023$+$0038 \citep{Stappers14,Takata13,Patruno14} and PSR J1824$-$2452I \citep{Papitto13} have directly shown transitions from LMXB to MSP states and vice versa. 
Such systems in tight orbits ($\lesssim$ 1 day) around main-sequence-like companions (with typical mass 0.2--0.3 $M_\odot$) are called 
``redbacks'' \citep{Mallory11}, similar to another class of eclipsing systems, called ``black-widows'', 
containing MSPs that are ablating their very-low-mass ($< 0.05 M_\odot$) companions. 
The discovery of more such LMXB--MSP systems will help to determine whether they are on the way to becoming canonical MSPs with white dwarf companions, 
or whether they are a sub-class of MSPs that will continue to transition back-and-forth between the two states.

PSR J1023$+$0038 was the first such system seen to transition from an LMXB to an eclipsing binary radio MSP \citep{Archibald09}.
PSR J1023$+$0038 disappeared as an observable radio MSP sometime in 2013 June \citep{Stappers14} suggesting a return to the LMXB state. \cite{Patruno14} reported that an accretion disc 
has recently formed in the system and also reported the detection of fast X-ray changes spanning about two orders of magnitude in luminosity. 
PSR J1023$+$0038 has also brightened by a factor of $\sim$5 in gamma-rays since the radio pulsations disappeared \citep{Stappers14}. The only other known gamma-ray 
emitting LMXB is XSS J12270$-$4859, which is positionally coincident with the \textit{Fermi} gamma-ray source 1FGL J1227.9$-$4852. Based on this coincidence and
similarities with PSR J1023$+$0038, XSS J12270$-$4859 was expected to be another system capable of transitioning to a radio MSP state \citep{Hill11}.

A sudden decrease of optical and X-ray brightness for XSS J12270$-$4859 between 2012 November 14 and 2012 December 21 was reported by \cite{Bassa14}.
The optical flux decreased by 1.5 to 2
magnitudes, while the Swift/XRT count rate decreased by at least a factor of 10 over the same time-period. The previous signs
of an accretion disc (i.e. double-peaked optical emission lines) also disappeared. {\it Chandra} and {\it XMM-Newton} observations presented by 
\cite{Bogdanov14} showed a (previously absent) orbital modulation of the X-ray brightness which, in analogy to PSR J1023$+$0038, is believed to 
come from an active pulsar wind shocking near the surface of the companion star \citep{Bogdanov11}. 
The XSS J12270$-$4859 observations suggested a transition from an LMXB-like state in 2012 to one where the accretion disc was absent in 2013. 
A faint non-thermal radio source was identified in the field of XSS J12270$-$4859 \citep{Hill11,Masetti06}. 
However, the flat spectral index of this emission seen during the LMXB state was probably caused by an outflow,
rather than radio pulsar emission, which is also reported for PSR J1023$+$0038 \citep{Deller14}. \cite{Bassa14} had
searched for radio pulsations using the Parkes at 1.4\,GHz, but pulsations were not found in a blind search.
This Letter describes the radio pulsar discovery with the GMRT and the follow-up study of PSR J1227$-$4853, the redback MSP associated with XSS~J12270$-$4859.

\section{Radio observations and search analysis}  
\label{sec:obs_analysis}

We searched for radio millisecond pulsations in XSS J12270$-$4859 with the 
Giant Metrewave Radio Telescope (GMRT) at 607\,MHz.  The GMRT Software Back-end \citep[\texttt{GSB};][]{Roy10} produces
simultaneous incoherent and coherent beam filter-bank outputs of 512$\times$0.0651 MHz channels at 61.44 $\upmu$s.
We performed sensitive coherent array observations using 70\% of the GMRT array 
with phase center at R.A. = 12$^\mathrm{h}$27$^\mathrm{m}$58\fs8, Decl. = $-$48\degr53\arcmin42\farcs1 \citep{Masetti06}. We recorded 3 1-hour 
long scans, beginning on 2014 Feb 12 at 20:46:15 UTC, interleaving with phase calibrator scans once an hour. 

We processed the data in 6 half-hour segments on the NCRA \texttt{IBM} cluster with Fourier-based acceleration search methods 
implemented in \texttt{PRESTO} \citep{Ransom02}.  We investigated trial dispersion measures
(DMs) ranging from 0 pc cm$^{-3}$ up-to 500 pc cm$^{-3}$. The Fourier-based periodicity search was done using harmonic
summing up-to 8 harmonics. 
Considering 32 MHz bandwidth, 10\% duty cycle, coherent array gain of $\sim$ 7 K/Jy,
for a 30-minute observing time, we estimate the search sensitivity as (92K $+T_\mathrm{sky}$)/(1020K) mJy for a 5-$\sigma$
detection at 607\,MHz. Considering $T_\mathrm{sky}$ $\sim$ 10K \citep{Haslam81}, our search sensitivity is $\sim$ 0.1 mJy.

In the first 30-minute chunk, we discovered a highly-accelerated 1.69 ms pulsar at a DM of 43.4 pc cm$^{-3}$ (reported in \citealt{Roy14}). 

\section{Follow-up radio timing} 
\label{sec:timing}

\begin{figure}[htb]
\includegraphics[width=3.4in]{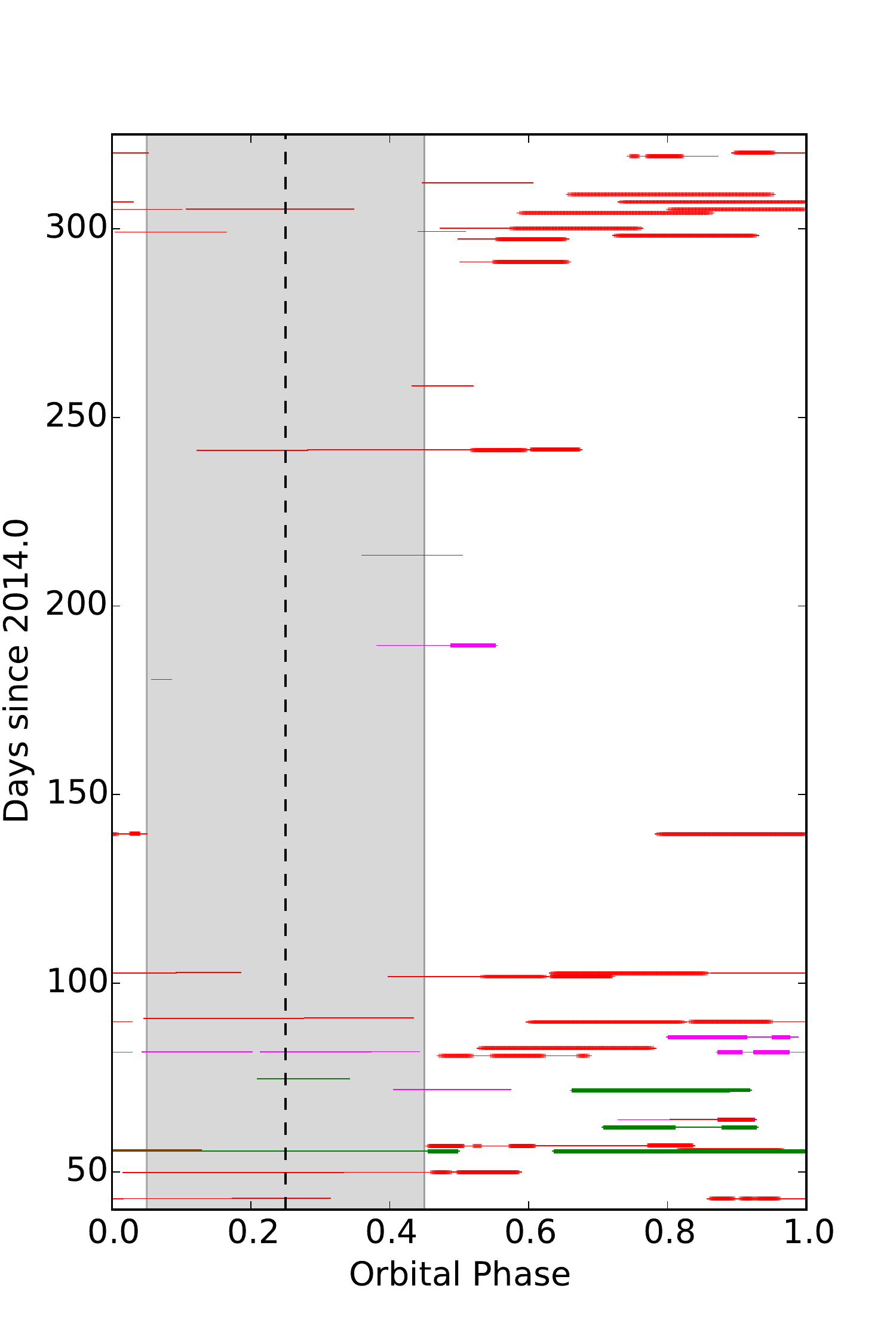}
\includegraphics[width=3.3in]{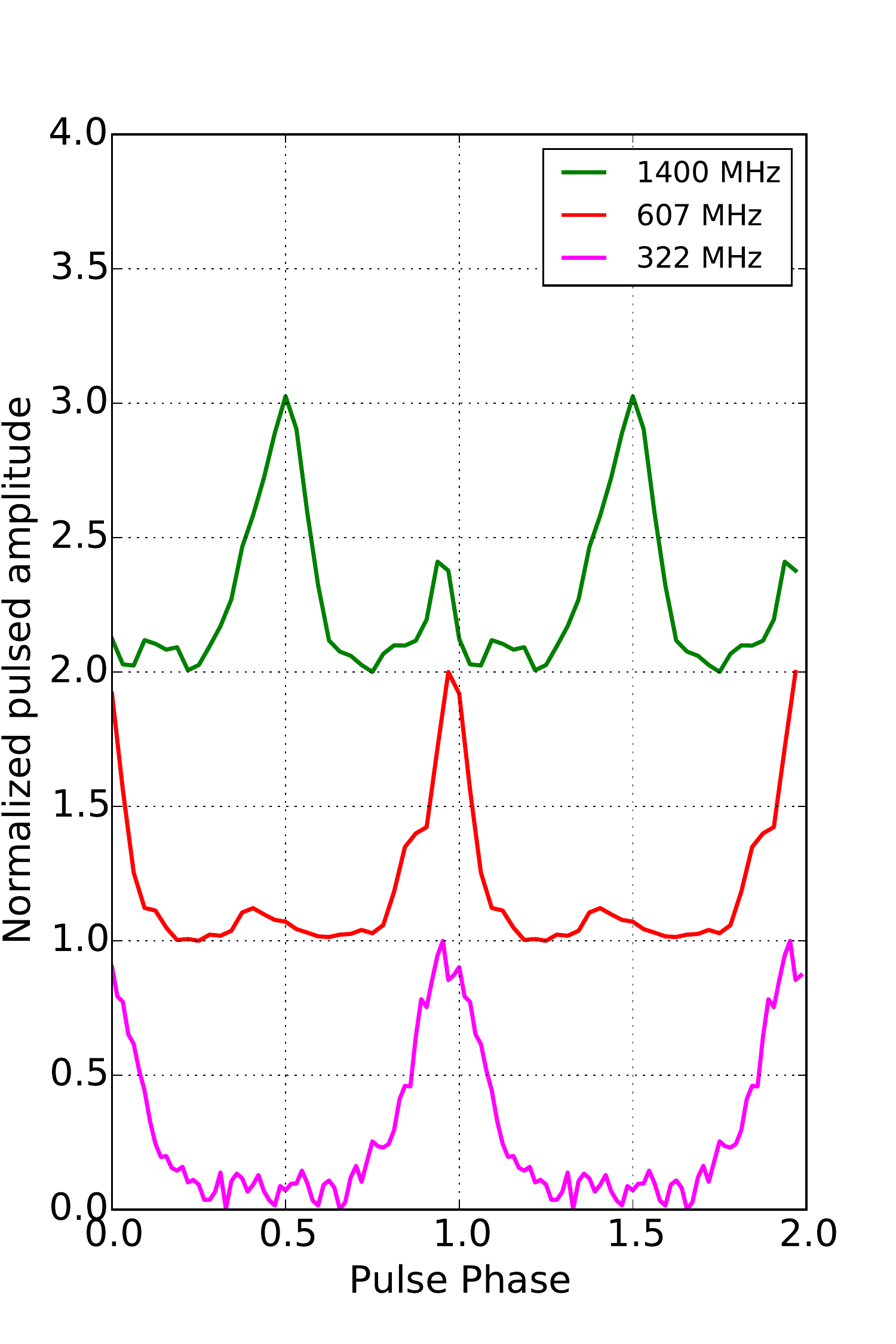}
\caption{The left panel shows the orbital phase coverage, where thick lines indicates detections. Red and magenta symbols correspond to GMRT observations at 607\,MHz and 
322\,MHz respectively and green for Parkes observations at 1.4\,GHz. The eclipse region is shaded and the superior conjunction is marked by the dotted line. 
The right panel shows the pulse profiles at three observing frequencies with vertical offsets added for clarity. GMRT-Parkes time-offset and the profile evolution can make the 
profile alignment covariant with DM.\label{fig:profile}}
\end{figure}

Following the discovery we started a timing campaign for PSR J1227$-$4853 using the GMRT coherent array at 607\,MHz. We also have a few detections of this pulsar at 322\,MHz 
using the GMRT and at 1.4\,GHz using the Parkes telescope. The timing observations span 270 days, from 2014 Feb 20 to 2014 Nov 17, consisting of 25 epochs, 
each typically 1-hour long. The filter-bank settings defined in Section \ref{sec:obs_analysis} were used for both the GMRT bands. 
The Parkes observations used the \texttt{DFB3} pulsar backend producing 512$\times$0.5 MHz channels at 80 $\upmu$s. The left panel of Fig.~\ref{fig:profile} shows the orbital phases of our 
observations.
The pulsar is not detected between orbital phases of 0.05 and 0.45 (phase 0.25 is superior conjunction of the pulsar) and 
there are also frequent intervals of non-detections well outside the main eclipsing region, which is quite similar to what is seen in PSR J1023$+$0038 \citep{Archibald13}.
We used multi-Gaussian fits to high signal-to-noise observations at each frequency to produce templates that were used for extracting time-of-arrivals (TOAs). 
Typically 10\,TOAs were derived from each observing session.

\begin{figure}[htb] 
\includegraphics[width=3.3in]{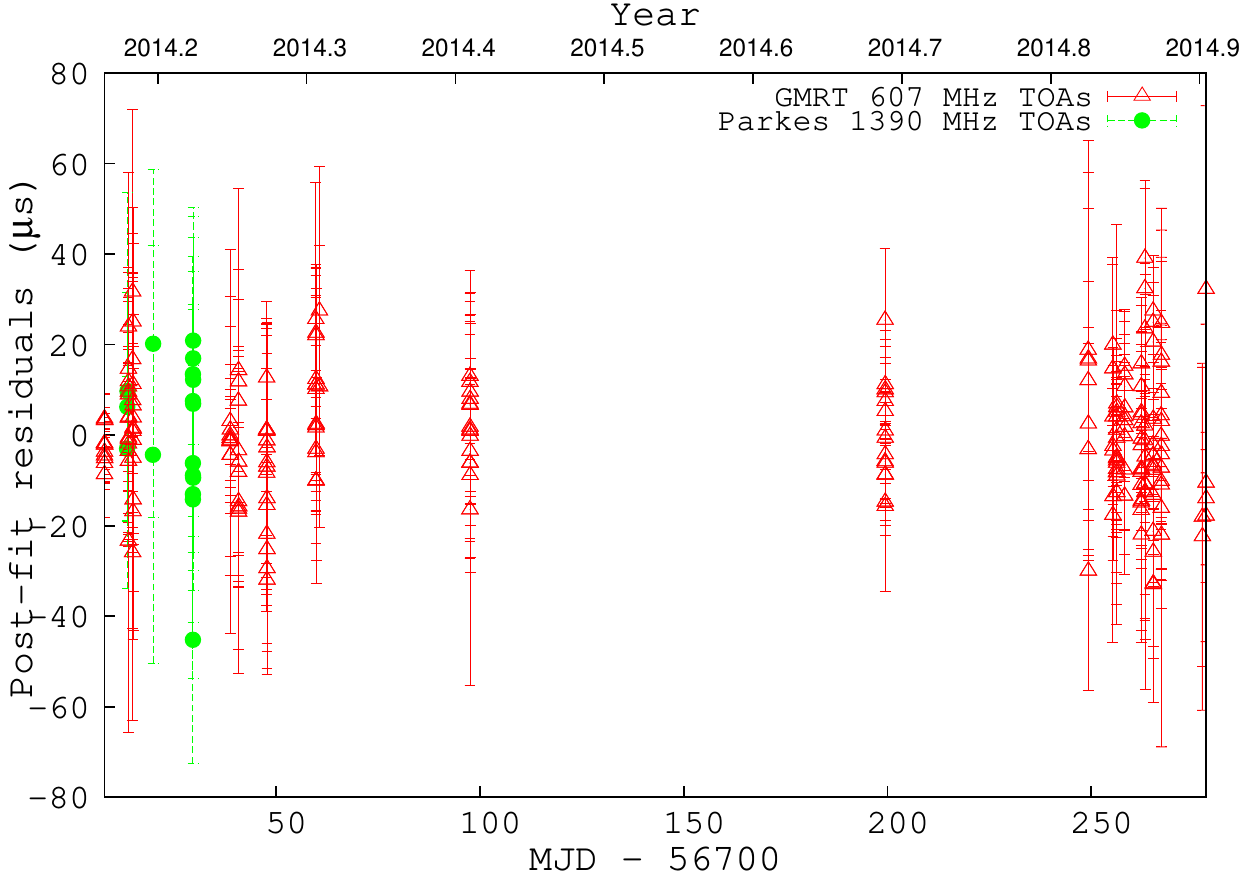}
\includegraphics[width=3.3in]{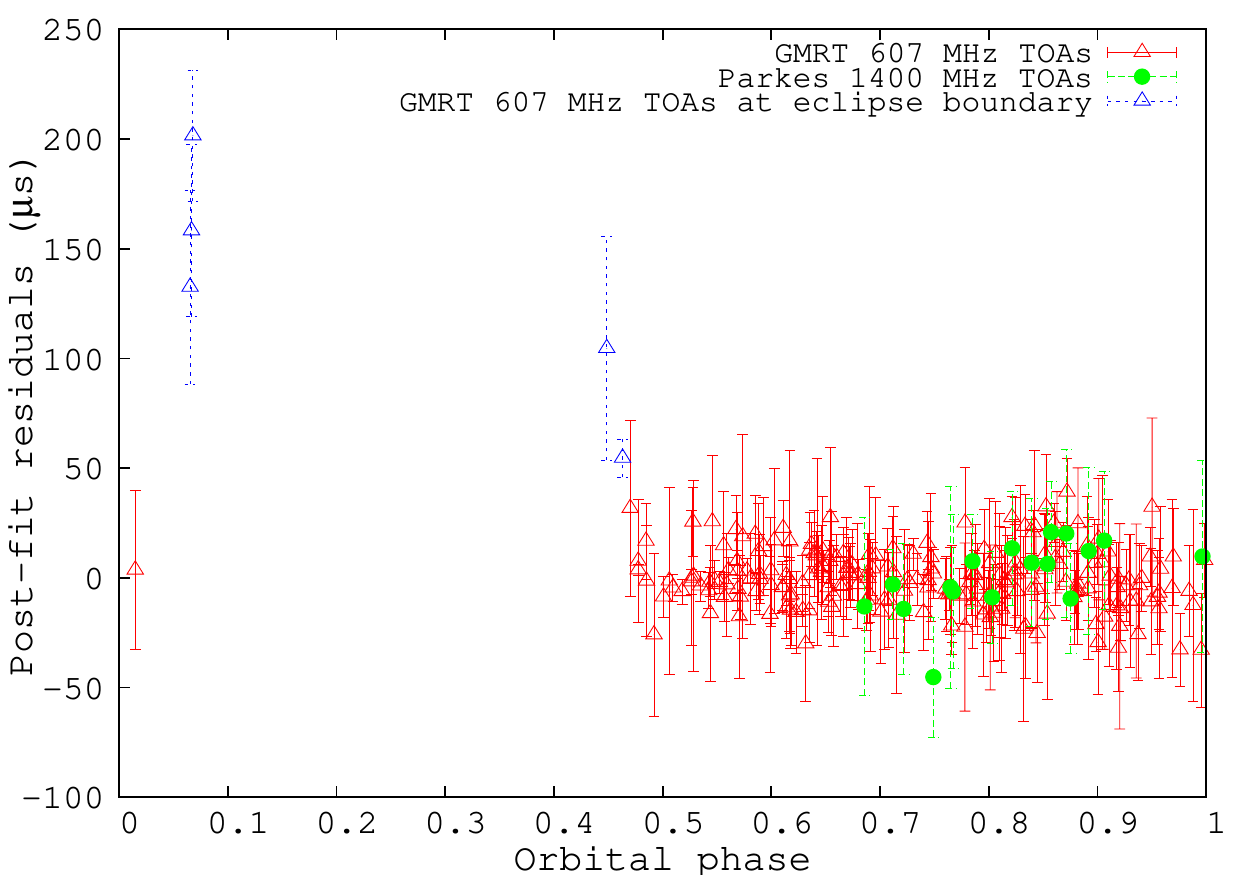}
\caption{Post-fit timing residuals of PSR J1227$-$4853 as a function of MJD (left panel) and binary phase (right panel). The large 100 days gap in the TOAs is caused 
by the observations at eclipsing binary phase on 56838 and 56871 MJD. The 607\,MHz TOAs are marked in red open-triangle and 1390\,MHz TOAs are marked in green circle; 
607\,MHz TOAs at the eclipse boundary are included only in the right panel plot 
(to illustrate the excess delay), but not in the timing fit. \label{fig:residual}}
\end{figure}

Using the 2MASS catalog position of R.A. = 12$^\mathrm{h}$27$^\mathrm{m}$58\fs748, Decl. = $-48$\degr53\arcmin42\farcs88 (J2000) \citep{Cutri03,Masetti06},
for the \textit{a priori} astrometric model and the JPL DE405 solar system ephemeris \citep{Standish04}, we obtained a phase-connected timing model 
using \texttt{TEMPO}\footnote{\url{http://www.atnf.csiro.au/research/pulsar/tempo}}.  We used a constant offset between the Parkes and the GMRT TOAs of 0.542 ms, 
which was measured using the global timing campaign data of PSR J1713$+$0747 taken simultaneously at a similar observing frequency \citep{Dolch14}. The global timing 
ephemeris having precise DM measured for that epoch was used to estimate this \texttt{JUMP}, which absorbs any fixed time offsets between the different backends and observatory clocks.  
The DM was determined using 607\,MHz GMRT TOAs and 1.4\,GHz Parkes TOAs excluding the orbital phases 0.0 to 0.5. 
The binary timing model used is ELL1 \citep{Lange01}. 
Since we do not measure a significant eccentricity (with an upper limit of $e$ $<$ $3.3\times 10^{-5}$), we set the eccentricity to zero in our timing model.
The best-fit timing model (given in Table~\ref{tab:params}) is obtained for 270 days of data excluding the TOAs around the main eclipse phase 
and achieved a post-fit rms timing residual of
10 $\upmu$s (Fig.~\ref{fig:residual}). 

The eclipse region is symmetrical around the orbital phase 0.25 with a duration of around 2.7 hours (40\% of the orbital period). The companion's Roche lobe radius
of 0.24 R$_{\odot}$  is much smaller than the opaque portion of the companion's orbit (4.9 R$_{\odot}$) indicating that ionized material lies far
outside the Roche lobe causing excess DM at the eclipse boundaries. The excess DM of 0.013 pc cm$^{-3}$ seen at the eclipse ingress (as seen in Fig. \ref{fig:residual})
corresponds to an added electron density of $> 4 \times 10^{16}$ cm$^{-2}$.

\begin{deluxetable}{ll}
\tabletypesize{\scriptsize}
\tablewidth{0pt}
\tablecaption{Parameters of PSR J1227$-$4853
\label{tab:params}}
\tablehead{
\colhead{Parameter} & \colhead{Value\tablenotemark{a}}
}
\startdata
  \hline
  \multicolumn{2}{c}{2MASS position\tablenotemark{b}} \\
  \hline
Right ascension (J2000)\dotfill & 12$^\mathrm{h}$27$^\mathrm{m}$58\fs748$\pm$0\fs06 \\
Declination (J2000)\dotfill     & -48\degr53\arcmin42\farcs88$\pm$1\farcs \\
  \hline
  \multicolumn{2}{c}{Parameters from radio timing} \\
  \hline
Right ascension (J2000)\dotfill & 12$^\mathrm{h}$27$^\mathrm{m}$58\fs724(1)\\
Declination (J2000)\dotfill     & -48\degr53\arcmin42\farcs741(3)\\
Pulsar frequency $f$ (Hz)\dotfill  & 592.987773605(4)\\
Frequency derivative $\dot{f}$ (Hz s$^{-1}$)\dotfill & $-$3.9 (4)$\times$10$^{-15}$ \\
Period epoch (MJD)\dotfill  & 56707.9764\\
Dispersion measure $\mbox{DM}$ (cm$^{-3}$ pc)\dotfill & 43.4235(7)\\
Binary model\dotfill & ELL1\\
Orbital period $P_{b}$ (days)\dotfill       & 0.287887519(1) \\
Orbital period derivative $\dot P_{b}$\dotfill 	    & $-$8.7(1) $\times$ 10$^{-10}$\\
Projected semi-major axis $x$ (lt-s)\dotfill & 0.668468(4) \\
Epoch of ascending node passage $T_{\rm {ASC}}$ (MJD)\dotfill & 56700.9070772(2)\\
Span of timing data (MJD)\dotfill & 56707.95--56978.13 \\
Number of TOAs\dotfill & 236 \\
Post-fit residual rms ($\upmu$s)\dotfill & 10.1\\
TOA error scale factor\tablenotemark{c}\dotfill & 2.5\\
Reduced chi-square\dotfill & 0.58 \\
 \hline
  \multicolumn{2}{c}{Derived parameters} \\
  \hline
Mass function $f$ (M$_{\odot}$)\dotfill  & 0.00386973\\
DM distance\tablenotemark{d} (kpc)\dotfill & 1.4\\
Flux density at 607\,MHz\tablenotemark{e} (mJy)\dotfill & 6.6(2)\\
Surface magnetic field $B_{s}$ ($10^{8}$ G)\dotfill & 1.36(6) \\
Spin down luminosity  \.{E} (10$^{35}$ erg s$^{-1}$)\dotfill & 0.90(8) \\
Characteristic age $\tau$ (Gyr)\dotfill & 2.4(2) \\
\enddata
\tablenotetext{a}{Errors in the last digit are in parentheses}
\tablenotetext{b}{\cite{Cutri03}}
\tablenotetext{c}{TOA uncertainties were multiplied by \texttt{EFAC} to correct for generally underestimated uncertainty produced by \texttt{TEMPO}.}
\tablenotetext{d}{\cite{Cordes02}}
\tablenotetext{e}{Measured with continuum imaging. Differs from the preliminary estimate of \cite{Roy14}, likely due to an un-calibrated coherent array gain and 
possible scintillation effects.}
\end{deluxetable} 

\section{Radio imaging and flux variations} 
\label{sec:imaging}

The GMRT interferometric visibility measurements allowed us to carry out continuum imaging of
the field of PSR J1227$-$4853. We used the observation on 2014 February 20
taken at 607\,MHz for imaging studies, where the pulsar was tracked from
rise-to-set resulting in 3.5 hours of on-source time recorded in multiple scans. In this time interval
the pulsed emission is seen only for the last 43 minutes of the observation, corresponding to orbital phases of 0.48 to 0.57.
The raw data were converted into \texttt{FITS} format, after which they were processed by
the flagging and calibration pipeline \texttt{flagcal}
\citep{Prasad12,Chengalur13}. Flux, phase and bandpass calibration were done
using nearby calibrator, J1154$-$3505. The
interpolated flux at 607~MHz was computed to be 7.9~Jy using the VLA calibrator
manual. All further processing was done using \texttt{AIPS}\footnote{\url{http://www.aips.nrao.edu}}.

\begin{figure}[htb]
\includegraphics[width=2.9in]{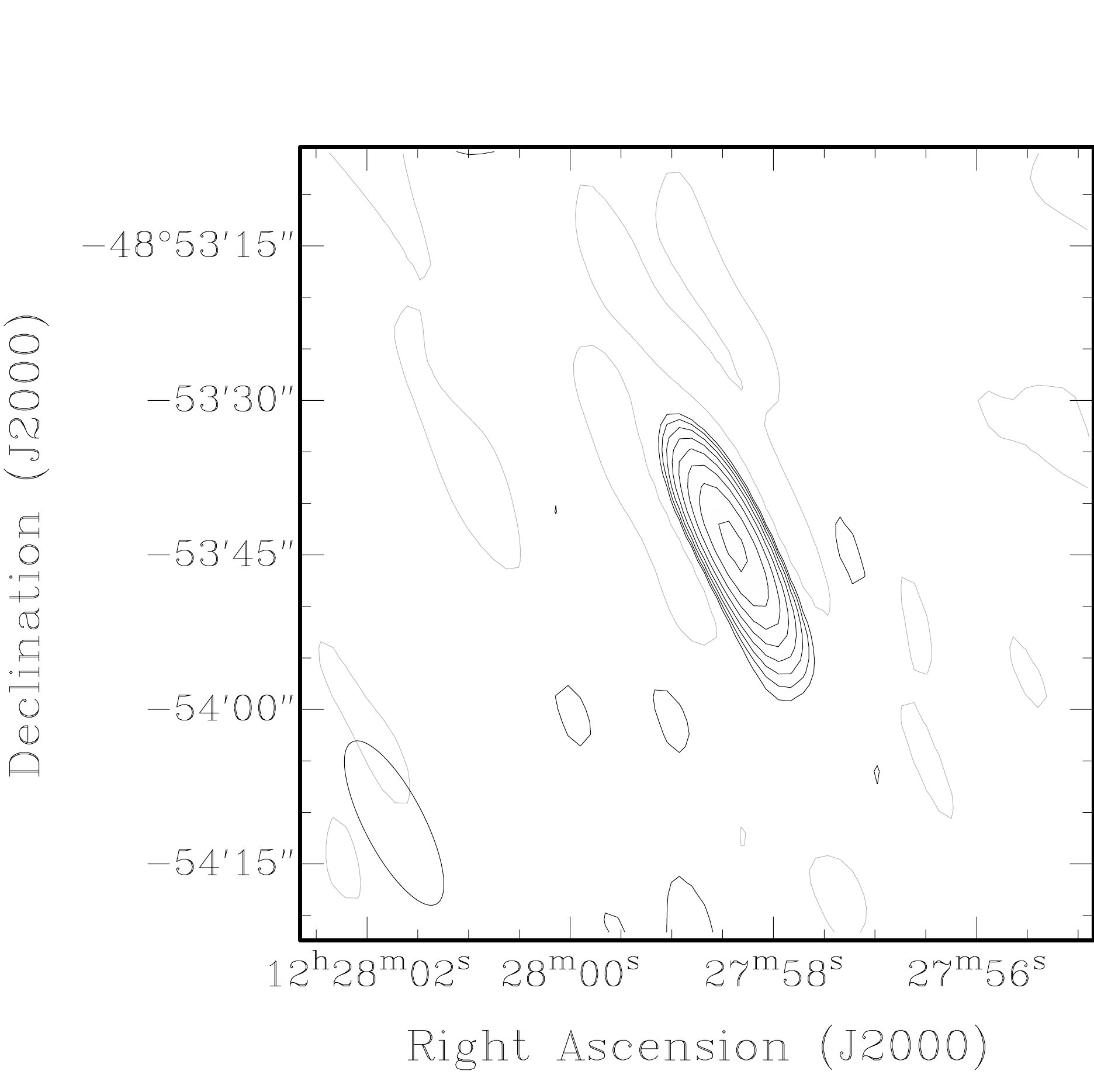}
\includegraphics[width=3.9in]{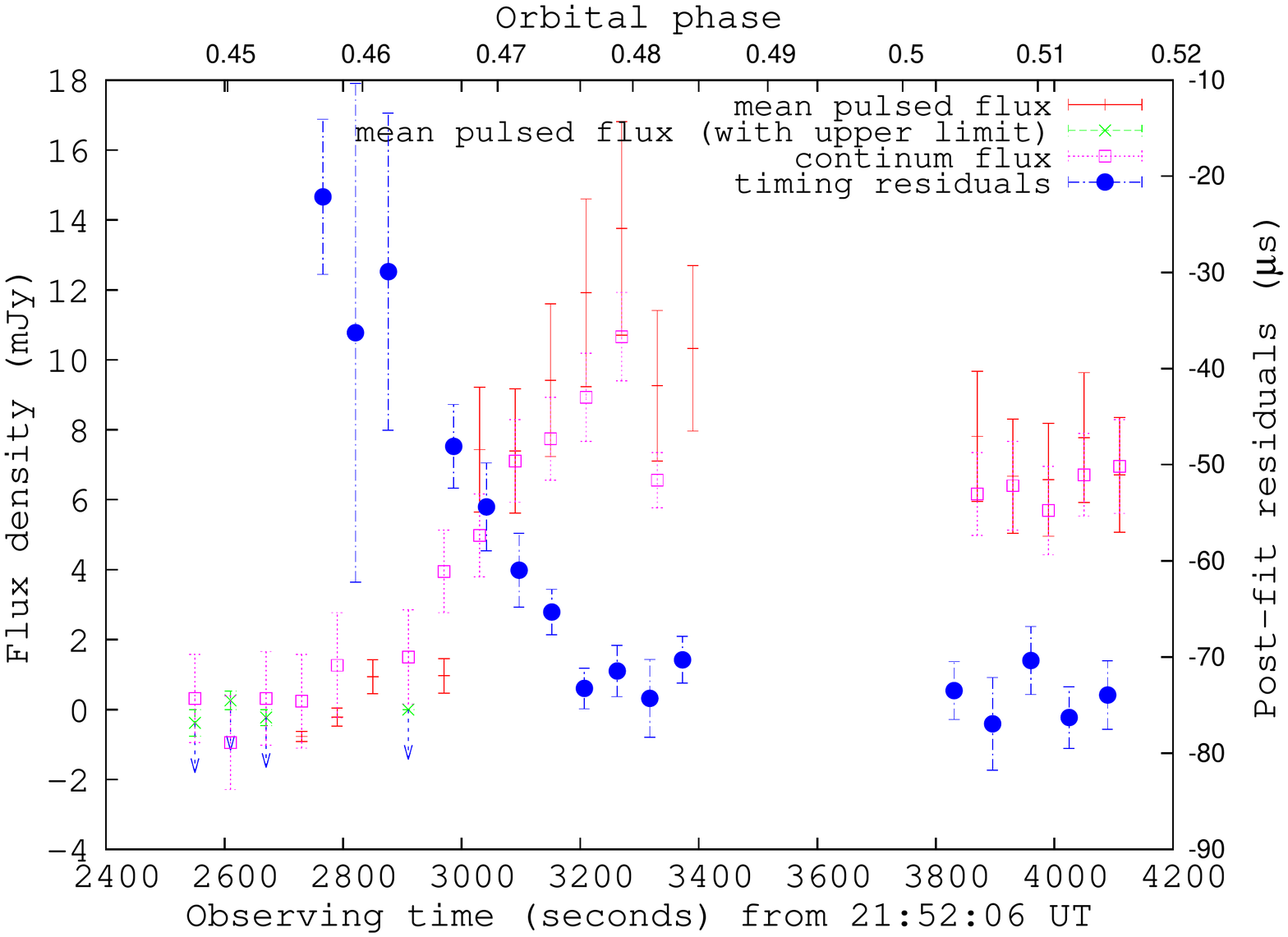}
\caption{Left panel: Radio emission at 607\,MHz at the location of PSR J1227$-$4853 from the GMRT observations over the orbital phase of 0.48 to 0.57. 
Contours start at 0.5 mJy/beam and increase in steps of $\sqrt{2}$. The 2-$\sigma$ negative contour is shown in grey. The synthesised beam (lower left corner) is
15\arcsec$\times$5\arcsec. Right panel: Variation of 
flux density at the eclipse egress. Continuum fluxes are plotted in purple and mean pulsed fluxes are plotted in red. Upper limits in mean 
pulsed flux density are given in case of non-detections. Timing residuals (right-side y-axis) at the eclipse egress is plotted in blue. 
There is a gap in the data between 3400--3800 seconds due to scan boundaries. \label{fig:flux}}
\end{figure}

Self-calibration was done 
using 30\arcsec\ $\times$ 21\arcsec\ resolution faceted images of the field made using the \texttt{3D} 
imaging option of \texttt{IMAGR}. A full resolution image of only the region around the target source was
made, separately for each of the scans.
Emission in the region of the target source
was seen only in the last scan (image in Fig. \ref{fig:flux}), which corresponds to the time interval where 
the pulsed emission is seen in the synchronously taken TOAs. The position and peak flux measured from
Gaussian fitting to the source are RA (2000) $=$ 12$^\mathrm{h}$27$^\mathrm{m}$58\fs4$\pm$0\fs3,
Decl. (2000) $=$ $-$48\degr53\arcmin44\farcs3$\pm$2\farcs5, and 6.6 $\pm$ 0.2~mJy respectively.
A light curve for this source was then determined by making images at
1-minute resolution. Fig. \ref{fig:flux} compares the continuum flux variation with the mean pulsed flux and the pulse arrival time delays as a function of orbital phase.  

During the main eclipse phase no emission is seen at the pulsar location. To
place the best possible limit, all of the detected sources, except from the
central facet, were subtracted from the visibility data. A further round
of self-calibration was done using only the sources in the central facet,
providing a better correction for the ionospheric phases. The final rms noise that we achieve is  85\,$\upmu$Jy
using the data for the full eclipse duration. \cite{Hill11} reported the detection
of a continuum source (at an epoch which corresponds to the X-ray accreting phase)
at the location of the pulsar with a flux of 190 $\upmu$Jy at 9~GHz and
a spectral index of $-0.1 \pm 0.1$ (based on reanalysis by \citealt{Bassa14}). Though our non-detection is consistent with the nominal spectral index, 
the emission observed by \cite{Hill11} could be powered by an outflow, which is absent in the GMRT image, as the outflow driven radio emission turns off during
radio MSP state \citep{Bassa14}.
 
\section{Pre-discovery radio observations}
\label{sec:pre-discovery}

Prior to the discovery of PSR J1227$-$4853, several searches for radio pulsations have been attempted using Parkes and GMRT. 
All of them (listed in Table \ref{tab:obs}) resulted in non-detections. 
We compute the pulsed flux density limits using nominal telescope parameters such as 
gain, bandwidth etc (\citealt{Ray11}, Table 19).
We consider a pulse duty cycle of 10\% and a 5-$\sigma$ detection threshold. Parkes 1390\,MHz on 2012 Mar 22 and GMRT 322\,MHz on 2012 Jul 23 are the only two observations 
spanning the non-eclipsing binary phases of the pulsar during the LMXB state. There is a marginal 5-$\sigma$ detection of pulsation for the 2012 Mar 22 Parkes observations, 
when folded with the radio timing ephemeris. We also folded the Parkes data from \cite{Bassa14} and detected the pulsar 
in several intervals of the 2013 Nov 13 observation (details in Table \ref{tab:obs}) at 1390\,MHz, at a significance below the threshold used for their blind search.  
\begin{deluxetable}{cccccc}
\tabletypesize{\scriptsize}
\tablewidth{0pt}
\tablecaption{Pre-discovery radio pulsation search observations 
\label{tab:obs}}
\tablenum{2}
\tablehead{\colhead{Telescope} & \colhead{Frequency } & \colhead{Date} & \colhead{$t_{obs}$} & \colhead{Orbital phase} & \colhead{$S_{min}$\tablenotemark{a}} \\ 
\colhead{} & \colhead{(MHz)} & \colhead{} & \colhead{(h)} & \colhead{} & \colhead{(mJy)} } 
\startdata
Parkes & 1390 & 2009 Nov 25\tablenotemark{b} & 2 & 0.12--0.41 & 0.15 \\
Parkes & 1390 & 2010 Jul 18\tablenotemark{b} & 1.1 & 0.05--0.21 & 0.20 \\
Parkes & 1390 & 2010 Nov 12\tablenotemark{b} & 1 & 0.25--0.40 & 0.21 \\
Parkes & 1390 & 2012 Mar 22 & 1 & 0.78--0.92 & 0.21 \\
GMRT\tablenotemark{c} & 322 & 2012 Jul 23 & 1 & 0.88--0.02 & 0.09 \\
Parkes & 1390 & 2012 Nov 07 & 0.75 & 0.16--0.2 & 0.25 \\
Parkes & 1390 & 2013 Nov 13\tablenotemark{d} & 1 & 0.43--0.57 & 0.20 \\
Parkes & 1390 & 2013 Nov 13\tablenotemark{d}{$_*$} & 1 & 0.58--0.72 & 0.20 \\
Parkes & 1390 & 2013 Nov 13\tablenotemark{d}{$_*$} & 1 & 0.73--0.87 & 0.20 \\
Parkes & 1390 & 2013 Nov 13\tablenotemark{d}{$_*$} & 1 & 0.87--0.02 & 0.20 \\
Parkes & 1390 & 2013 Nov 13\tablenotemark{d} & 1 & 0.02--0.16 & 0.20 \\
Parkes & 1390 & 2013 Nov 17\tablenotemark{d} & 1 & 0.48--0.62 & 0.20 \\
Parkes & 1390 & 2014 Jan 09\tablenotemark{d} & 1 & 0.79--0.93 & 0.20 \\
Parkes & 3100 & 2014 Jan 09\tablenotemark{d} & 0.58& 0.96--0.04 & 0.54 \\
Parkes & 732 &  2014 Jan 09\tablenotemark{d} & 0.58& 0.96--0.04 & 0.19 \\
\enddata
\vspace{0.3cm}\\
\tablenotetext{a}{$S_{min}$ is quoted for 607\,MHz using spectral index of $-$1.7}
\tablenotetext{b}{\cite{Hill11}}
\tablenotetext{c}{Incoherent array mode}
\tablenotetext{d}{\cite{Bassa14}, $*$ with detections}
Back-ends used for Parkes observations: up-to 2012 using \texttt{AFB}; 2013 and 2014 observations at 1390 MHz using \texttt{BPSR}; \texttt{DFB3} and \texttt{DFB4} 
for 732 MHz and 3100 MHz respectively\\
\end{deluxetable}

\section{Discussion} 
\label{sec:discussion}

We report the GMRT discovery of a redback MSP, PSR J1227$-$4853, associated with the LMXB system XSS J12270$-$4859, confirming that the system now hosts an active radio MSP
as predicted by \cite{Bassa14}. This is the third system after PSR J1023$+$0038 and PSR J1824$-$2452I showing evidence of state switching between LMXB and MSP states.
We report results from radio timing observations with the GMRT and Parkes spanning over 270 days from 2014 Feb 20 to 2014 Nov 17.  Comparing TOAs derived from 
the 2013 Nov 13 Parkes observation from \cite{Bassa14} indicates $\sim$ 2\,seconds change in $T_\mathrm{ASC}$ relative to the extrapolation of our timing model. This is
similar to orbital phase changes seen in PSR J1023$+$0038 by \citet{Archibald13}. A model with up-to third-order orbital period derivatives is required to accommodate 
such a rapid change, similar to what is observed in PSR J2051$-$0827 \citep{Lazaridis11}.

Based on our radio timing, we measure $P_\mathrm{b}$ = 0.287887519(1) days and obtain the first
detection of $\dot P_{b} = -8.7(1) \times 10^{-10}$.
Though the $P_\mathrm{b}$ determined from radio timing is consistent at the 2-$\sigma$ level with
the photometric period of 0.28804(8) days from \cite{Bassa14}, our measurement differs by 0.000132(2) days from the spectroscopic
$P_\mathrm{b}$ measurement of 0.2880195(22) days by \cite{deMartino14}  obtained in 2012 March/April and 2013 December. This
difference can not be explained by the observed $\dot P_{b}$. Instead, it appears \cite{deMartino14} miscounted
the number of orbital revolutions between their observations. The zero-crossing time ($T_0$)
from \cite{deMartino14} show that their spectroscopic $P_\mathrm{b}$ counts 2172 orbital revolutions,
one revolution less when using the more accurate and correct $P_\mathrm{b}$ derived from pulsar timing.

Converting the $T0$ measurements by \cite{deMartino14} from HJD to TDB we find that the radio timing ephemeris provides orbital phases ($\phi$) 
of 0.244(6) and 0.263(17) for the 2012 March/April and 2013 December epochs. These correspond to inferior conjunction
of the companion star, which is consistent with the $T_\mathrm{ASC}$ used in the timing ephemeris. 
Our timing ephemeris, yields an $\phi$ of 0.997(7) for the zero phase of the photometric lightcurve given in \cite{Bassa14}. 
Note that time-definition in \cite{Bassa14} is in error; it is TDB instead of HJD (Cees Bassa private communication).
In conclusion, the optical phase measurements by \cite{Bassa14} and \cite{deMartino14} are
consistent with our timing ephemeris of a pulsar orbiting an irradiated binary companion.

The timing ephemeris also provides the projected radial velocity amplitude of the pulsar $K_\mathrm{1}=50.622619464(1)$\,km\,s$^{-1}$. Combined with the
projected radial velocity amplitude of the companion determined from optical spectroscopy, $K_\mathrm{2}=261(5)$\,km\,s$^{-1}$ \citep{deMartino14}, this constrains 
the mass ratio to $q=M_2/M_1=0.194(3)$. 
For an edge-on orbit, the measurement of the mass-ratio and the mass-functions set
1-$\sigma$ lower limits on the mass of the companion and the pulsar as 0.142\,M$_\odot$ and 0.719\,M$_\odot$, respectively. Further
constraints are provided by limiting the orbital inclination to $i\ga43\degr$ \citep{deMartino14}. 
The absence of X-ray eclipses leads these authors to suggest $i\la73\degr$. For this range in inclination the companion
mass is constrained to 0.167 to 0.462\,M$_\odot$, while the pulsar has a mass of 0.86 to 2.38\,M$_\odot$.
The combination of $q$ and $K_\mathrm{2}$, allows an estimate of
the rotational broadening of spectral lines of the companion through 
$v_\mathrm{rot} \sin i=0.46 K_2 [(1+q)^2q]^{1/3}=78(12)$\,km\,s$^{-1}$. This implicitly assumes that the
companion is filling its Roche lobe and is tidally locked with the binary orbit \citep{Wade88}. This prediction of
the rotational broadening is consistent with the measured value of $v_\mathrm{rot}\sin i=86(20)$\,km\,s$^{-1}$ during the LMXB state 
\citep{deMartino14} (though the rotational broadening is close to the instrumental resolution), suggesting that the binary 
companion of PSR J1227$-$4853 was (or was close to) filling its Roche lobe prior to the 2012 transition. Modeling of the light curves for 
companion stars of other redbacks has also shown them to be (near) Roche lobe filling \citep{Breton13}. 

We also report a spin-down luminosity, $\dot E$ = 0.9 $\times$ 10$^{35}$ erg s$^{-1}$, which is $\sim$ 10$^{3}$ times the X-ray luminosity in the MSP-state.  
This places PSR J1227$-$4853 among the few highly energetic MSPs with $\dot E$ $\sim$ 10$^{35}$ erg s$^{-1}$. 
The Shklovskii correction for $\dot P$ considering a typical MSP velocity (85 km\,s$^{-1}$; \citealt{Toscano99}) is $<$ 1\%. 
The measured $\dot E$/$a^2$ ($=$ $\sim$ 4.3 $\times$10$^{33}$ erg lt-s$^{-2}$s$^{-1}$), where $a$ ($=$ 1.97 R$_{\odot}$) is the separation between the pulsar and companion,  
is indicative of greater energy flux of the system (and is similar to that of PSR J1023$+$0038) to efficiently 
drive off material from the surface of the companion. 

The simultaneous timing and imaging studies of PSR J1227$-$4853 provide a powerful probe of the eclipse mechanism. The very well correlated variation of continuum 
flux density and mean pulsed flux density at the eclipse boundary helps to rule out effects such as excess dispersion or 
scattering being the cause of eclipse. Rather, cyclotron-synchrotron absorption of the radio waves by electrons is likely to be the cause of the radio eclipse \citep{Thompson94}. 

Since the discovery of PSR J1227$-$4853 we have radio timing observations at the GMRT spanning 270\,days, where the MSP is consistently bright. Thus it is clear 
that the pulsar has transitioned from an accreting state around the end of 2012 and is now in a radio MSP state for at least 362 days (including the 2013 Nov 13 detections). 
The marginal detection of radio pulsations in 2012 may suggest that the system had entered 
a propeller state possibly alternating with a rotational-powered state on short timescales \citep{Patruno13}. 
The discovery of radio pulsations in XSS J12270$-$4859 further establishes that the sudden decrease in X-ray and optical flux (as well as the disappearance of 
double-peaked emission lines) relates to a cessation of accretion in the system, which has now firmly transitioned to a radio MSP state.
Considering the recent reverse swing from MSP to LMXB state in PSR J1023$+$0038, PSR J1227$-$4853 may also return back to the LMXB state within the coming year(s), 
resulting in the disappearance of radio pulsations. Then it will be possible to study what type of accretion state the system enters and whether material is being 
channeled onto the neutron star (\citealt{Archibald14}; \citealt{Papitto14}).

\acknowledgments

The GMRT is run by the National Centre for Radio Astrophysics of the Tata Institute of Fundamental Research, India. 
We acknowledge support of GMRT telescope operators for observations. We thank Andrew Lyne for discussion on the GMRT timing model.
The Parkes radio telescope is funded by the Commonwealth of Australia for operation as
a National Facility managed by CSIRO. We thank the help of John Reynolds in understanding the time offset at 
GMRT while combining with Parkes data. This work at NRL was supported by the Chief of Naval Research (CNR).
B.B. acknowledges support of Marie Curie grant (FP7) of EU. J.W.T.H. acknowledges funding from an NWO Vidi fellowship and ERC Starting Grant ``DRAGNET'' (337062).


\end{document}